\documentclass[12pt,a4paper,showkeys]{revtex4}
\usepackage{graphicx}
\usepackage{epsfig}
\usepackage{bm}
\usepackage{amsmath}
\usepackage{amssymb}
\usepackage{graphics}
\usepackage{epstopdf}
\usepackage[linkcolor=red,citecolor=green,urlcolor=blue]{hyperref}

\begin{document}

\title{ \textbf{QCD analysis of forward neutron production in DIS}}
\author{Federico Alberto Ceccopieri}
\email{federico.alberto.ceccopieri@cern.ch}
\affiliation{IFPA, Universit\'e de Li\`ege,  All\'ee du 6 ao\^ut, B\^at B5a,   \\4000
Li\`ege, Belgium}
\begin{abstract}
\noindent
We consider forward neutron production in DIS within fracture functions formalism.
By performing a QCD analysis of available data we extract proton-to-neutron fracture functions
exploiting a method which is in close relation with the factorisation theorem for 
this class of processes.
\end{abstract}

\maketitle

\section{Introduction}
\label{intro}
\noindent 
In hadronic collisions a portion of the produced particle spectrum 
is characterised by hadrons carrying a sizeable fraction of the available energy and produced at small polar angle with respect to the collision axis. It is phenomenologically observed that for such hadrons their valence-parton composition is almost or totally conserved
with respect to the one of inital-state hadrons~\cite{Basile}. Such semi-inclusive processes are instrumental to study the scaling hypothesis of forward hadron production cross sections~\cite{Feynman_scaling,limiting_pp} and give insight on non-perturbative aspects of QCD dynamics in high energy collisions.
These hadrons, in fact, are produced at very small transverse momenta with respect to the collision axis, a regime where perturbative techniques can not be applied. 
Quite interestingly, forward particle production has also been observed in processes which involve point-like probes in lepton-hadron interactions, such as Semi-Inclusive Deep Inelastic Scattering (SIDIS).
At variance with the hadronic collisions mentioned above, such process involves a large momentum transfer at the lepton vertex. The presence of a hard scale is a basic requirement in the derivation of a dedicated factorisation theorem~\cite{factorization_soft,factorization_coll} which ensures  that collinear QCD factorisation holds in the leading-twist approximation for forward particle production in DIS. 
The relevant cross sections can then be factorised into perturbatively calculable short-distance cross sections and new distributions, fracture functions, which
simulatenously encode informations both on the
interacting parton and on the non-perturbative QCD dynamics of the spectator fragmentation into the observed forward hadron. Despite of being non-perturbative in nature, their scale dependence can be calculated within perturbative QCD~\cite{trentadue_veneziano}.
Fracture functions obey in fact DGLAP~\cite{DGLAP} inhomogeneous evolution equations which result from the structure of collinear singularities 
in the target-fragmentation region~\cite{trentadue_veneziano,graudenz}.
The dedicated factorisation theorem~\cite{factorization_soft,factorization_coll} 
guarantees that fracture functions are universal distributions, at least in the context of SIDIS. 
On this theoretical basis, an impressive experimental program has been pursued at HERA in diffractive DIS which led to accurate determination of the so called diffractive parton 
distributions, \textsl{i.e.} proton-to-proton fracture functions in the very forward limit,
allowing for the first time a quite accurate investigation of the parton content of the pomeron. 
The whole formalism has been later used in Ref.~\cite{ceccopieri_mancusi} to 
extract proton-to-Lambda fracture functions within a combined perturbative QCD fit to available SIDIS data.

In this paper we will focus on forward neutron production in DIS which 
provides, with respect to the aforementioned processes, complementary informations on soft QCD dynamics. 
An intensive physics program with forward neutron tagging has been pursued at HERA as well, where  recent results~\cite{H1_14,ZEUS_07} show that around 8\% of the DIS events contain a forward neutron.
These data are crucial in testing the limiting fragmentation hypothesis~\cite{limiting_DIS} and have been used as benchmark in a number of Regge-based models
~\cite{Bishari,Holtmann,Kaidalov,Kopeliovich,nikolaev,pirner,khoze} which mainly concentrate on the modelisation of forward neutron production mechanisms. 
In the present paper we adopt instead a complementary approach and no modelisation 
of neutron production mechanisms is attempted. This strategy is in line with the factorisation
theorem. The resulting set of proton-to-neutron fracture functions (nFFs) 
could then be used in hard-scattering factorisation test~\cite{Klasen} 
in forward neutron tagged dijet photoproduction in $ep$ collisions,
as already measured at HERA~\cite{ZEUS_dijet_photo_plus_neutron,H1_dijet_photo_plus_neutron}, 
where factorisation is expected to hold only for the so-called direct
component of the cross section. Even more intriguing appears 
to be the possibility of using nFFs for predicting the cross section 
for the associated production of a forward neutron and a Drell-Yan pair
or dijet system in hadronic collisions. For this processes the factorisation theorem
is not expected to hold and therefore nFFs determined by DIS data alone offer the opportunity 
to gauge factorisation breaking effects. 

The paper is organised as follows. In Sec.~\ref{data_obs}
we describe the process under study, the kinematic variables relevant
to the analysis and the observable used in the fit. In Sec.~\ref{theory}
we discuss the evolution of fracture functions and the general method 
with which we build initial conditions for the QCD evolution. 
In Sec.~\ref{ph1} we describe the details of the QCD fit and in Sec.~\ref{ph2} we 
assess the impact of experimental and theoretical errors on the obtained 
neutron FF set. In Sec.~\ref{conclusions} we summarise our results. 

\section{Data set and observable}
\label{data_obs}
\noindent
In this analysis we consider semi-inclusive DIS events of the type 
\begin{equation}
\label{process}
e^+ (k) \,+ \,p(P ) \,\rightarrow \,  e^+ (k^{'} )\, + \,n(P_n )\, + \, X(p_X )\,,
\end{equation}
where, beside the outgoing lepton, an additional neutron $n$ is detected in the final state. 
In eq.~(\ref{process}) $X$ stands for the unobserved part of the hadronic system and 
particles four-momenta  are indicated in parenthesis. 
The kinematic variables $Q^2$, $x_B$ and $y$ are used to describe the inclusive DIS scattering
process. They are defined as
\begin{equation}
q=k-k', \;\; Q^2=-q^2, \;\; x_B = \frac{Q^2}{2P \cdot q}, \;\; y=\frac{P \cdot q}{P \cdot k}\,.
\end{equation}
The kinematic variables used to describe the final state neutron are the 
neutron transverse momentum $p_T$ evaluated with respect to the beam axis
and the longitudinal momentum fraction $x_L$ defined by
\begin{equation}
x_L=1-\frac{q \cdot (P-P_n)}{P \cdot q} \simeq E_n/E_p\,,
\end{equation}
where $E_n$ and $E_p$ are the neutron and proton energy in the laboratory frame, respectively.  
In the following we use the scaled fractional momentum variable $\beta$ 
defined by
\begin{equation}
\beta=\frac{x_B}{1-x_L}\,,
\end{equation}
where $1-x_L$ is the maximum available fractional momentum of the parton partecipating 
the hard scattering.
The analysis is performed on the H1 $ep$ data of Ref.~\cite{H1_10} with positrons and protons energies respectively of $E_e=27.6$ GeV and $E_p=920$ GeV, corresponding 
to a centre-of-mass energy of $\sqrt{s}=319$ GeV.
The kinematic range of the selected DIS events is $6 < Q^2 < 100$ Ge$V^2$ , $0.02 < y < 0.6$ 
and $1.5 \cdot 10^{-4} < x_B < 3 \cdot 10^{-2}$.
The values of $x_L$ range from $0.365$ to $0.905$. 
The kinematic $\beta$-coverage is $x_L$-dependent, in particular
$\beta_{min}=3.52 \cdot 10^{-4}$ at $x_L=0.365$  and $\beta_{max}=0.22$ at $x_L=0.905$.
Forward neutron production is characterised by small values of $p_T$. 
In Ref.~\cite{H1_10} an upper limit on $p_T$ is used to define  
the semi-inclusive forward neutron cross section which is correspondingly 
integrated up to $p_{T,max}=0.2$ GeV. Data are presented as a three-fold reduced $e^+p$ cross section, $\sigma_r^{LN(3)}$, which depends on the leading neutron transverse and longitudinal structure functions $F_2^{LN(3)}$ and $F_L^{LN(3)}$, respectively. In the one-photon exchange approximation, it reads:
\begin{equation}
\label{sigmar}
\sigma_r^{LN(3)}(\beta,Q^2,x_L)=F_2^{LN(3)}(\beta,Q^2,x_L)-\frac{y^2}{1+(1-y)^2} F_L^{LN(3)}(\beta,Q^2,x_L)\,.
\end{equation} 

\section{Theory setup}
\label{theory}
\noindent
Hard-scattering factorisation for this class of processes states that the 
structure functions in eq.~(\ref{sigmar}) are of the form 
\begin{equation}
\label{hard_fact}
F_k^{LN(3)}(\beta,Q^2,x_L,p_T^2)=\sum_i \int_{\beta}^1
\frac{d\xi}{\xi} \; M_{i/P}^N(\beta,\mu_F^2;x_L,p_T^2) \; C_{ki} 
\bigg( \frac{\beta}{\xi},\frac{Q^2}{\mu_F^2},\alpha_s(\mu_R^2)\bigg)
+\mathcal{O}\bigg( \frac{1}{Q^2}\bigg)\,.
\end{equation}
The index $i$ runs on the flavour of the interacting parton.
The hard-scattering coefficients $C_ {ki}$ ($k=2,L$) are pertubatively calculable 
as a power expansion in the strong coupling $\alpha_s$ and 
depend upon $\mu_F^2$ and $\mu_R^2$, the factorisation and renormalisation scales,
respectively. The $C_{ki}$ coefficient functions are the same as in fully inclusive DIS.
The proton-to-neutron fracture functions $M_{i/P}^N(\beta,\mu_F^2,x_L,p_T^2)$ can be interpreted as 
the number density of interacting partons at a scale $\mu_F^2$ and fractional momentum $\beta$ conditional to the observation of a forward neutron 
in the final state specified by a fractional momentum $x_L$ and transverse momentum squared $p_T^2$.
They contain non-perturbative informations on the fragmentation of the spectator system which results from the hard interactions. 
The $p_T$-unintegrated nFFs appearing in eq.~(\ref{hard_fact}) obey 
standard DGLAP~\cite{DGLAP} evolution equations~\cite{extM}. In the case $p_T$ is integrated over up to values of order $Q^2$, 
neutron fracture functions obey an inhomogenous DGLAP-type evolution equations~\cite{trentadue_veneziano}. The additional inhomogeneous term accounts for neutron production coming from the fragmentation of initial state parton radiation. In the present case, where the $p_T$ of the neutron is integrated up to some $p_{T,max}$ which lies in the non-perturbative region, neutron fracture functions are defined as
\begin{equation}
\label{intM}
M_{i/P}^N(\beta,Q^2,x_L)=\int^{p_{T,max}^2} d p_T^2 \, M_{i/P}^N(\beta,Q^2,x_L,p_T^2)
\end{equation}
and again obey familiar DGLAP evolution equations~\cite{CT_fracture_ISR}
\begin{equation}
\label{Mevo}
Q^2 \frac{\partial M_{i/P}^N(\beta,Q^2,x_L)}{ \partial Q^2}=
\frac{\alpha_s(Q^2)}{2\pi}\int_{\beta}^1 \frac{du}{u} \,
P_i^j(u) \, M_{j/P}^N\Big(\frac{\beta}{u},Q^2,x_L\Big)\,,
\end{equation}
valid at fixed values of $x_L$.
The central problem of this type of analyses is to find sensible initial 
conditions for the relevant distributions prior to evolution. 
One may resort to phenomenological models to describe forward neutron production. 
In general at low and intermediate $x_L$ the dominant mechanism is 
expected to be proton-remnant fragmentation into neutrons, while 
at high $x_L$ the exchange of virtual particles is expected to dominate. 
In the present analysis, we work at fixed $x_L$ and no attempts to model this non-perturbative 
dynamic at the proton vertex is made.
Since hard-scattering factorisation in the form eq.~(\ref{hard_fact}) holds at fixed values of $x_L$ and $p_T^2$ and this dependence is fully contained in nFFs, 
these conditional parton distributions are uniquely fixed by the kinematics of the outgoing neutron and they are, at least in principle, different for different values of $x_L$ and $p_T^2$. 
The approach we describe in this paper fully takes into account these important recipes in the construction of sensible input for nFFs distributions
focusing on parton dynamics as explored by the virtual
photon once an additional forward neutron is detected in the final state. 
This new approach have already been used in the extraction of diffractive parton densities from diffractive DIS data in Ref.~\cite{mywork_DPDF}.
This idea is realised in practice performing a series of QCD fits 
at fixed values of $x_L$ with a common initial condition 
controlled by a set of parameters $\{p_i\}$. 
This procedure guide us to infere the approximate dependence of parameters 
$\{p_i\}$ on $x_L$ allowing the construction of a generalised 
initial condition in the $(\beta,x_L)$-space to be used in a $x_L$-combined QCD fit. It is important to note that if four-differential
cross sections were available, the same method could be used to test 
whether, at fixed $x_L$, the parton content explored by the virtual photon is the same (a part normalisation) in different neutron $p_T$ ranges.

\section{Fitting procedure}
\label{ph1}
\noindent
In this section we describe QCD fits at fixed values of $x_L$.
The distributions of neutron FFs in the quark sector at large 
$\beta$ may show valence-like structures for some quark-flavour combinations.
However the accessible values of $\beta$ in the experimental data are quite low.
In view of this fact, and in order to reduce the number of free parameters, we assume 
that all light quark distributions are equal to each other, so that only the singlet and gluon distributions are required. We assume for the latter, at the arbitrary scale $Q_0^2$ and for any given value of $x_L$, a momentum distributions of the type:
\begin{eqnarray}
\label{ic_singlet}
\beta \, M_{\Sigma/P}^N(\beta,Q_0^2) &=& A_q \, \beta^{B_q} \, (1-\beta)^{C_q} \,,\\
\label{ic_gluon}
\beta \, M_{g/P}^N(\beta,Q_0^2) &=& A_g \, \beta^{B_g} \, (1-\beta)^{C_g}\,. 
\end{eqnarray}
These distributions are then evolved with the \texttt{QCDNUM17}~\cite{QCDNUM17} package  
within a zero mass (ZM) variable flavour number scheme (VFNS) to next-to-leading order accuracy.
Within this scheme heavy flavours ditributions are generated radiatively above their respective mass thresholds which are set to $m_c=1.4$ GeV and $m_b=4.5$ GeV for charm and 
bottom quark, respectively. In the VFNS the initial conditions must be 
imposed at $Q_0^2<m_c^2$. 
The factorisation and renormalisation scale are both set equal to $Q^2$. The coupling constant is set to $\alpha_s(M_Z^2)=0.118$.
The convolution engine of \texttt{QCDNUM17} is used to obtain 
$F_2^{LN(3)}$ and $F_L^{LN(3)}$ structure functions at next-to-leading order 
which are then used to calculate the reduced cross section in eq.~(\ref{sigmar}) and 
minimised against data by using the \texttt{MINUIT} program~\cite{MINUIT}. 
We adopt the generalised $\chi^2$ definition proposed in Ref.~\cite{PZ}
\begin{equation}
\chi^2=\sum_i \Bigg( \frac{m_i-f_i(\mbox{\bf{p}},\mbox{\bf{s}})}{\sigma_i}\Bigg)^2+\sum_k s_k^2\,,
\end{equation}
where systematics effects are incorporated in theory model predictions
\begin{equation}
f_i(\mbox{\bf{p}},\mbox{\bf{s}}) = t_i (\mbox{\bf{p}}) + \sum_k s_k \Delta_{ik}\,.
\end{equation}  
Here $m_i$ is the measurement of data point $i$, $t_i$ is the model prediction depending
on a set of parameters $\bf{p}$, $\sigma_i$ are the uncorrelated and statistical errors on data point $i$ added in quadrature and $\Delta_{ik}$ is the correlated systematic error from source $k$ on the data $i$. The variables $s_k$ denote Gaussian random variables with zero mean and unit variance. In the present Section we use the above definitions with $s_k=0$.  

No cut on the invariant mass of the hadronic system $X$ nor on the minimum $Q^2$ of data 
to be included in the fit is applied.
\begin{table}
\begin{center}
\begin{tabular}{c|c|c} \hline \hline
 \hspace{0.4cm} $x_L$      \hspace{0.4cm}  & \hspace{0.4cm} $\chi^2$   
\hspace{0.4cm} &  Fitted points \\ \hline
0.365  &  12.0  &  29 \\
0.455  &  25.5  &  29\\
0.545  &  19.9  &  29\\
0.635  &  21.0  &  29\\
0.725  &  23.6  &  29\\
0.815  &  17.1  &  29\\
0.905  &  15.7  &  29\\ \hline \hline
 Sum      &  134.8 & 203\\ \hline
\end{tabular}
\caption{\small{$\chi^2$ values for fits at fixed values of $x_L$ with initial 
condition in eqs.~(\ref{ic_singlet},\ref{ic_gluon}). The total $\chi^2$,
calculated as the sum of partial $\chi^2$ at fixed $x_L$, 
and the total number of fitted points are also indicated.}}
\label{fixed_xl-fits}
\end{center}
\end{table}
As already discussed above, given the kinematic coverage of the data, 
we found that the singlet large-$\beta$ coefficient $C_q$ is loosely constrained by data. For the gluon distribution, which is only indirectly constrained by scaling violations, we found that $B_g$ strongly correlates with $A_g$, when the 
former is left free to vary in the fit.
For these reasons we set temptatively $C_q=0.5$, $C_g=1$ and $B_g=0$ so that the initial condition contains three free parameters in each $x_L$-bin. 
We performed a combined scan on the value of initial scale $Q_0^2$. Given the quite 
stiff functional form of the initial conditions, there is a mild dependence of
the $\chi^2$ on $Q_0^2$ which is then fixed at $Q_0^2=1$ Ge$\mbox{V}^2$.
An essential condition for the $x_L$-combination procedure to work is that good quality fits must be obtained in each $x_L$-bin with the common initial conditions, eqs.~(\ref{ic_singlet},\ref{ic_gluon}). 
\begin{figure}[t]
\begin{center}
\includegraphics[width=8cm,height=6cm]{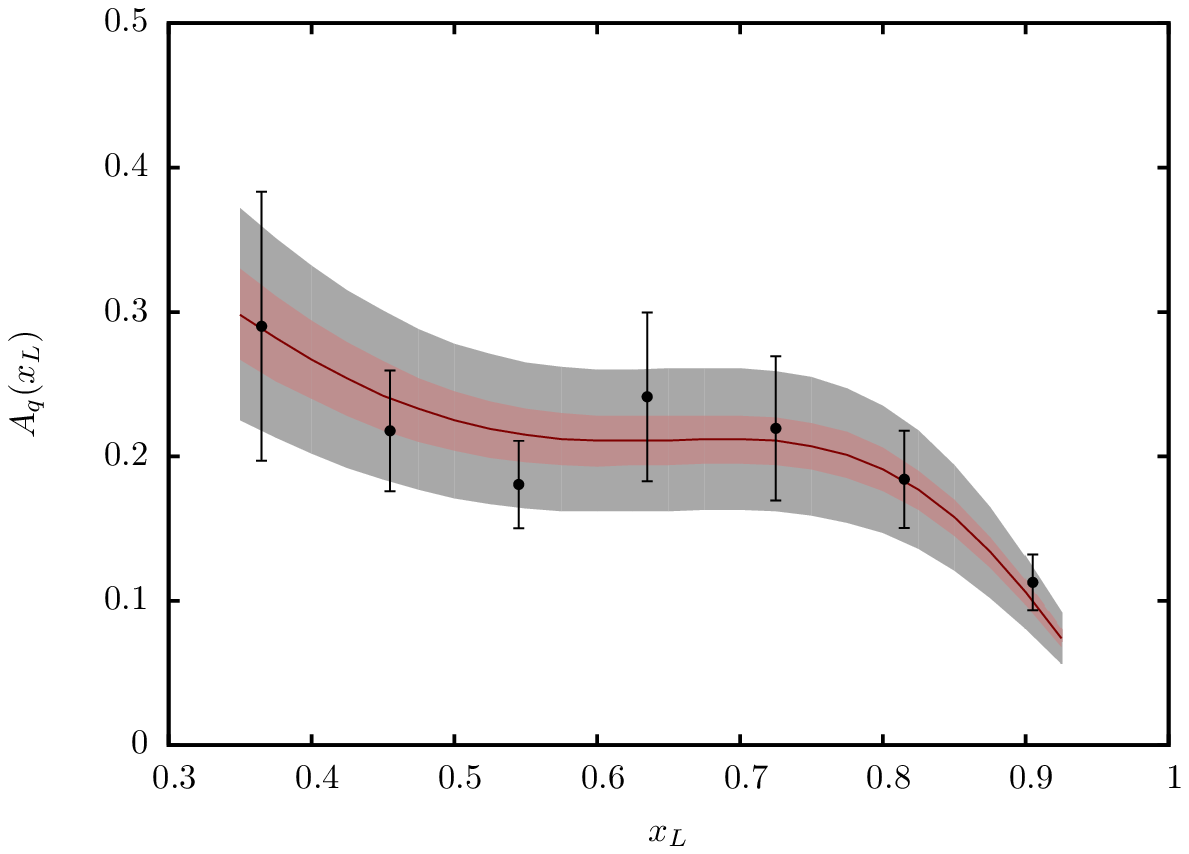}
\includegraphics[width=8cm,height=6cm]{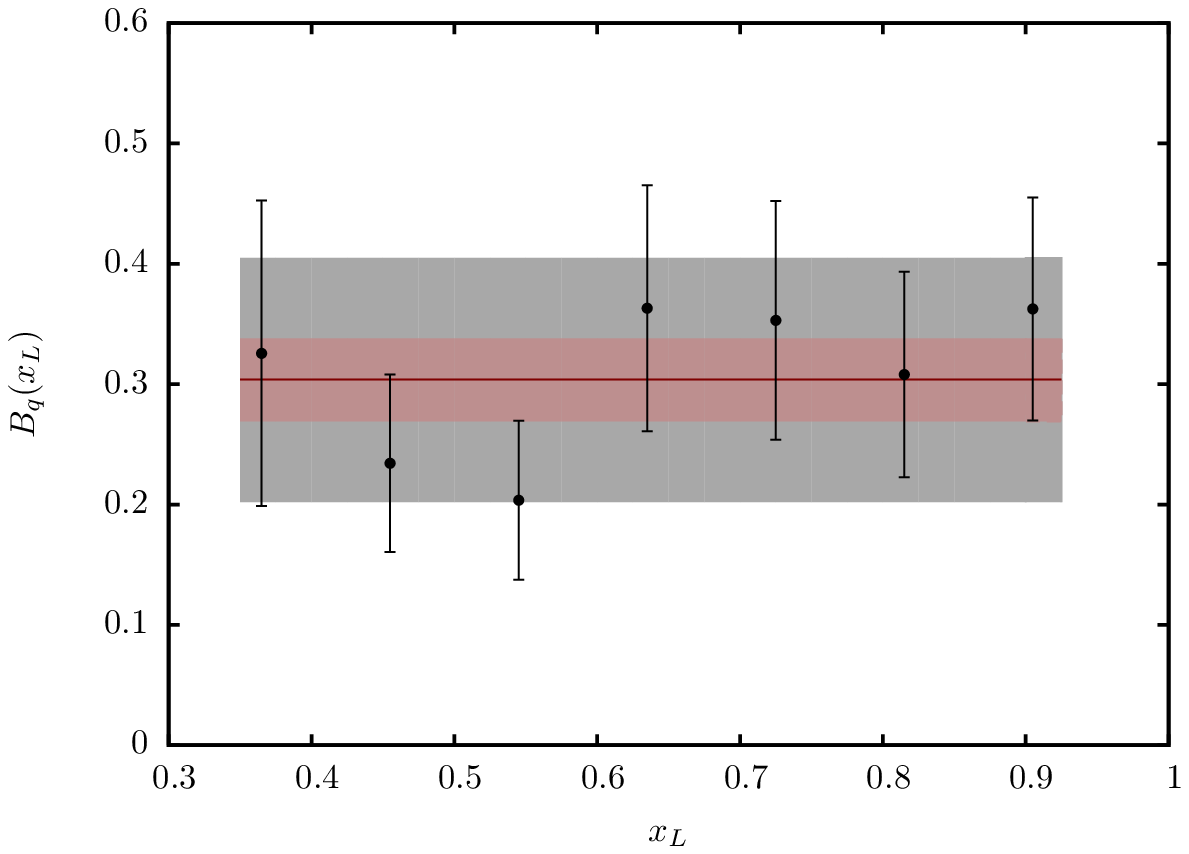}
\includegraphics[width=8cm,height=6cm]{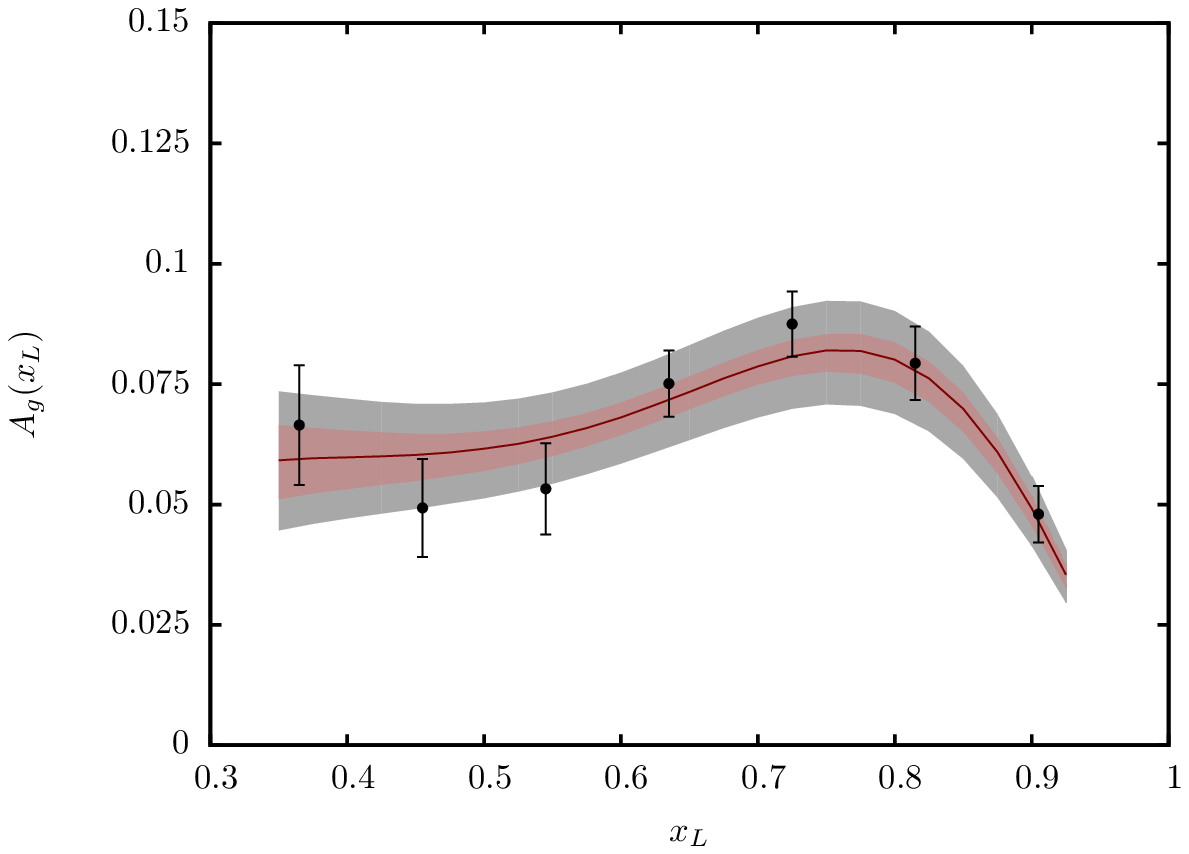}
\caption{The black points represent parameters $A_q$, $B_q$ and $A_g$ and their errors as a function of $x_L$, as obtained from fixed-$x_L$ fits. The best fit parametrisations according to eqs.~(\ref{coeffAq2},\ref{coeffBq2},\ref{coeffAg2}) is displayed as a black solid line. The light brown  and light grey bands are obtained by propagating statistical and uncorrelated errors in the $x_L$-combined fit, as described in the text, with the condition $\Delta \chi^2=1$ and $\Delta \chi^2=9$, respectively.}
\label{xL-dependence}
\end{center}
\end{figure}
The $\chi^2$ values of the fixed-$x_L$ fits, obtained with statistical and uncorrelated uncertainties added in quadrature, are presented in Tab.~(\ref{fixed_xl-fits}). From these values we may conclude that initial conditions provided by eqs.~(\ref{ic_singlet},\ref{ic_gluon}), supplemented by the constraints on $C_q,C_g$ and $B_g$, are general enough to describe the data in all $x_L$-bins. 
\begin{table}[t]
\begin{center}
\begin{tabular}{c|c|c} \hline \hline
 \hspace{0.4cm} $x_L$      \hspace{0.4cm}  & \hspace{0.4cm} $\chi^2$   
\hspace{0.4cm} &  $\Delta \chi^2$ \\ \hline
  0.3650 &  12.7 & +0.7\\
  0.4550 &  27.5 & +2.0\\
  0.5450 &  22.0 & +2.1\\
  0.6350 &  22.3 & +1.3\\
  0.7250 &  25.5 & +1.9\\
  0.8150 &  17.3 & +0.2\\
  0.9050 &  16.3 & +0.6\\ \hline \hline
    Tot  &  143.6 & +8.8 \\ \hline
\end{tabular}
\caption{\small{$\chi^2$ values for the combined-$x_L$ fit. In the third coloumn is indicated the $\chi^2$ difference between combined-$x_L$ and fixed-$x_L$ fits in each $x_L$-bin.
The total $\chi^2$ of the combined-$x_L$ fit and the combination penalty are also indicated.}}
\label{combo_xl-fit}
\end{center}
\end{table}
With these results at hand we may now proceed and discuss the $x_L$-combined fit. 
The generalised initial conditions have now the form 
\begin{eqnarray}
\beta \, M_{\Sigma/P}^N(\beta,Q_0^2,x_L) &=& A_q (x_L)\, \beta^{B_q(x_L)} \, (1-\beta)^{C_q(x_L)} \,,\nonumber\\
\beta \, M_{g/P}^N(\beta,Q_0^2,x_L) &=& A_g (x_L)\, \beta^{B_g(x_L)} \, (1-\beta)^{C_g(x_L)}\,,
\label{ic2}
\end{eqnarray}
where the $\beta$ dependence is the same as in eqs.~(\ref{ic_singlet},\ref{ic_gluon})
and $x_L$ dependence is accounted for by the coefficients. 
The dependences of the $A_q$, $B_q$ and $A_g$ free parameters on $x_L$ may be inferred 
inspecting Fig.~(\ref{xL-dependence}). 
We adopt a redundant parametrisation of the coefficients of the type
\begin{eqnarray}
\label{coeffAq}
      A_q(x_L)&=&a_1 \,x_L^{b_1}\, (1+c_1 \,x_L^{d_1}) (1-x_L)^{e_1}\,, \\
\label{coeffBq}
      B_q(x_L)&=&a_2+b_2 \, x_L	+ c_2 x_L^2\,,\\
\label{coeffAg}
      A_g(x_L)&=&a_3 \,x_L^{b_3} \, (1+c_3 \,x_L^{d_3}) (1-x_L)^{e_3}\,,
\end{eqnarray}
where the term $1+c \,x_L^{d}$ is included to describe the relative maximum of the
normalisation $A_q$ and $A_g$ at intermediate values of $x_L$. For $B_q$ we assumed a second order polynomial in $x_L$. The parameters which were fixed in the fixed-$x_L$ fits are kept fixed to same values. All the QCD settings are the same as in fixed-$x_L$ fits. 
With the help of eq.~(\ref{ic2}) and eqs.~(\ref{coeffAq},\ref{coeffBq},\ref{coeffAg})
we perform a series of $x_L$-combined fits. At each iteration we study the eigenvalues of the covariance matrix of the fit parameters. Small eigenvalues, in fact, are associated to (combination of) parameters which are loosely determined by data.
We found that, at large $x_L$, both the singlet and gluon normalisations can be described by a common parametrisation $g(x_L)$ 
\begin{equation}
g(x_L)=(1+c_1 \,x_L^{d_1}) (1-x_L)^{e_1}\,.
\end{equation} 
The $B_q$ coefficient is found to be compatible with a constant so that only the parameter
$a_2$ is left free to vary in the fit. Finally we found that $b_1$ is determined with rather large error and compatible with zero, so that we fix it to this value. The final form of the parametrisations of the coefficients is then 
\begin{eqnarray}
\label{coeffAq2}
      A_q(x_L)&=&a_1 \, g(x_L)\,, \\
\label{coeffBq2}
      B_q(x_L)&=&a_2\,,\\
\label{coeffAg2}
      A_g(x_L)&=&a_3 \, x_L^{b_3} \, g(x_L)\,,
\end{eqnarray}
\begin{table}[t]
\begin{center}
\begin{tabular}{c|c} \hline \hline
\hspace{0.4cm} Parameter \hspace{0.4cm}  & \hspace{0.5cm} $p_i$   
$\pm$  $\delta p_i$  \hspace{0.5cm} \\ \hline
     $a_1$  &        0.62   $\pm$    0.07  \\ 
     $c_1$  &       17.2    $\pm$    1.8   \\   
     $d_1$  &        6.25   $\pm$    0.17  \\    
     $e_1$  &        1.77   $\pm$    0.05  \\ 
     $a_2$  &        0.30   $\pm$    0.03  \\  
     $a_3$  &        0.32   $\pm$    0.06  \\  
     $b_3$  &        0.90   $\pm$    0.27  \\ \hline      
\end{tabular}
\caption{\small{Best Fit parameter values and their errors.}}
\label{Best-fit_pars}
\end{center}
\end{table}
\begin{figure}[t]
\begin{center}
\includegraphics[width=8cm,height=7cm]{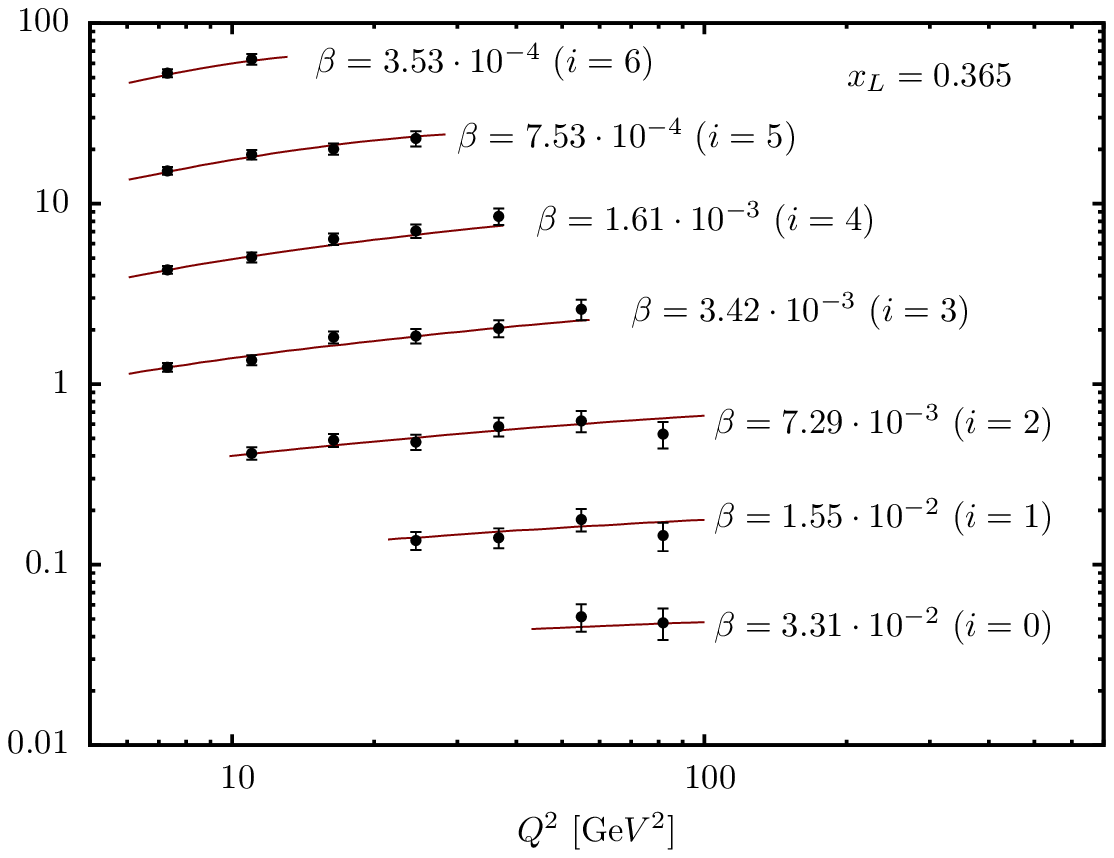}
\includegraphics[width=8cm,height=7cm]{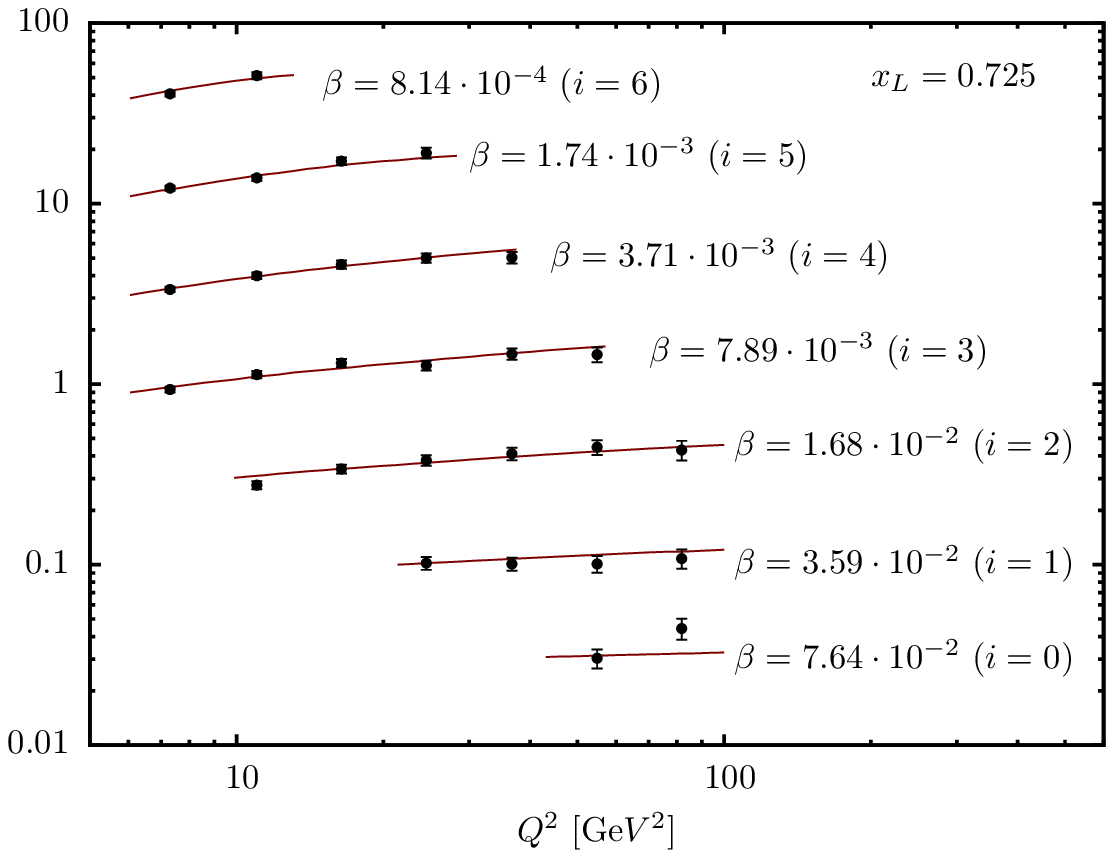}\\
\includegraphics[width=8cm,height=7cm]{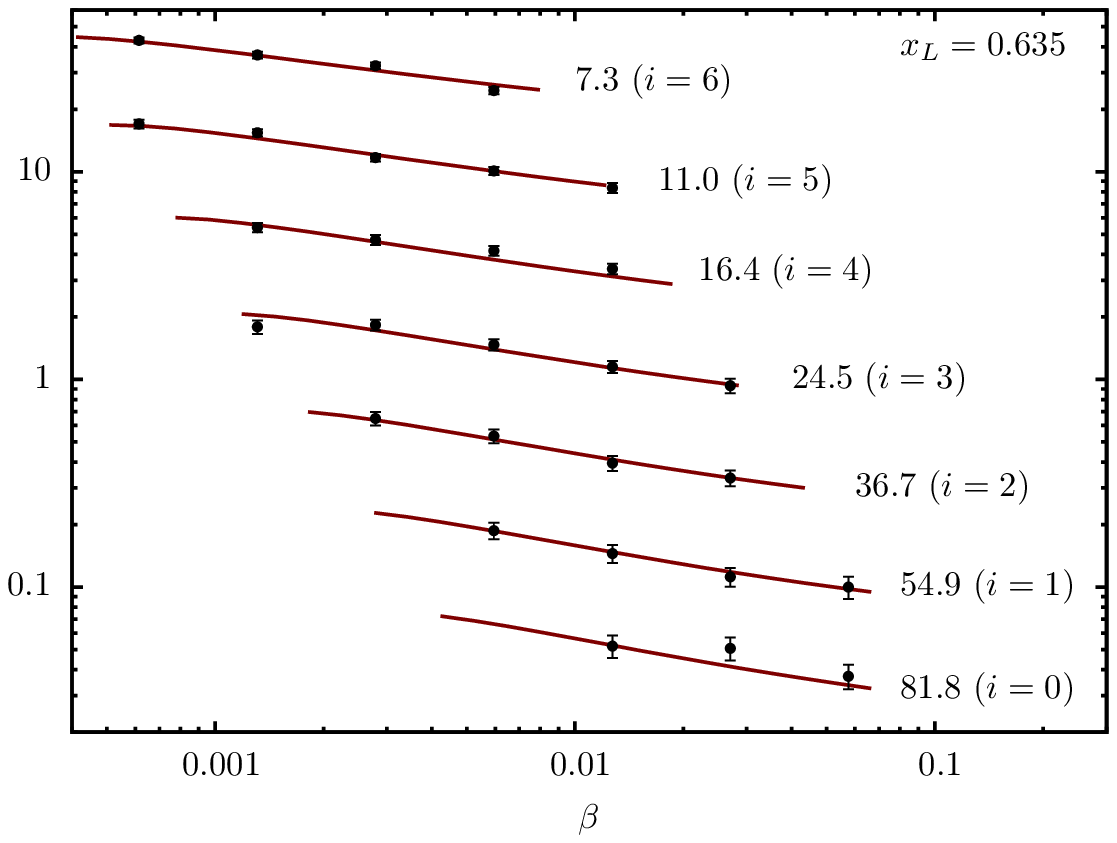}
\includegraphics[width=8cm,height=7cm]{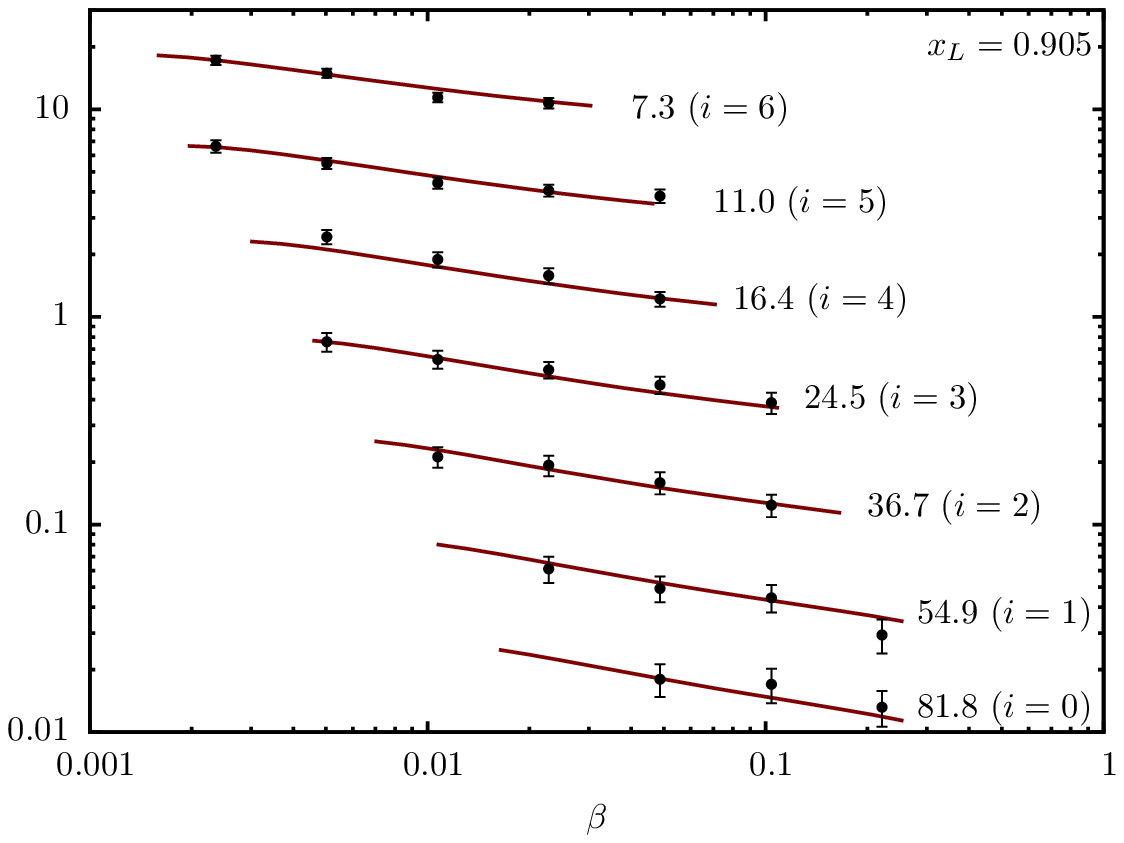}\\
\caption{Top: reduced cross section as a function of $Q^2$ at different $\beta$
in two bins of $x_L$. Bottom: reduced cross section as a function of $\beta$ at different $Q^2$
(in Ge$\mbox{V}^2$ units) in two bins of $x_L$ . The reduced cross section is scaled by a factor $3^i$ for better visibility. H1 points from Ref.~\cite{H1_10}. The error bars
associated with the data points show the sum in quadrature of the statistical and total systematic uncertainty.}
\label{sigmar_vs_q2}
\end{center}
\end{figure}
for a total of seven free parameters. The best fit, with statistical and uncorrelated errors added in quadrature, returns a value of 
$\chi^2$=143.6 for 196 degrees of freedom. The resulting $\chi^2$ in each 
$x_L$-bin is presented in Tab.~(\ref{combo_xl-fit}). In the third coloumn 
of the same table is indicated the increase in $\chi^2$ generated by 
the combination procedure in each $x_L$-bin with respect to the result 
presented for fixed-$x_L$ fit in Tab.~(\ref{fixed_xl-fits}).
Comparing the partial $\chi^2$ for each value of $x_L$ in Tab.~(\ref{combo_xl-fit}) and Tab.~(\ref{fixed_xl-fits}) we conclude that the combined-$x_L$ fit does not introduce any misrepresentation in the description of any given $x_L$-bin. We further quantify the quality 
of the choosen coefficient parametrisations comparing the sum of the $\chi^2$ obtained by
fixed-$x_L$ (134.8) with the $\chi^2$ from the combined fit (143.6). 
The combination procedure induces an increase in the overall $\chi^2$ of 8.8 units
and, on average, around one unit across the $x_L$ bins.
The best fit parameters and relative errors are reported in Tab.~(\ref{Best-fit_pars}).
The best fit predictions are compared to H1 data in Fig.~(\ref{sigmar_vs_q2})
in two representative $x_L=0.365$ and $x_L=0.725$ bins. 
From the plot it appears that the hard-scattering formula in eq.~(\ref{hard_fact}) together with nFF initial conditions describe data down to the lowest accessible value of $Q^2$.
This in turn implies also that, in the kinematical range covered by the experiment, no additional power-suppressed terms are required to describe the data. 
The presence of large and positive scale violations up to the largest 
parton fractional momentum, $\beta$, reveals the substantial contribution to $\sigma_r^{LN(3)}$ of the gluon nFF distribution induced by QCD evolution.
As a final remark we note that the singlet and gluon normalisation coefficients, $A_q$ and $A_g$, in eq.~(\ref{ic2}), have a different behaviour at small $x_L$. This in turn implies  
a violation of the so-called proton vertex factorisation. 
If this hypothesis is enforced, that is if we set $b_1=b_3$ and let this parameter free to vary in the fit, we obtain a $\chi^2=150$. Therefore, in the explored kinematical range and given the accuracy of the present data, proton vertex factorisation holds to a good approximation.
  
\section{Error estimation and propagation}
\label{ph2}
\noindent
In order to judge the agreement with other data sets and observables or to assess
effects beyond the ones taken into account by the theoretical model, the obtained nFF parametrisation must be supplemented, fully exploiting the potential of the data, by a careful error analysis. The general method with which experimental and theoretical uncertainties are propagated to a generic observable $F$ is based on the construction of alternative nFFs parametrisation sets $S_k$. By defining the difference $r_k=F(S_k)-F(S_0)$ and indicating with $S_0$ the best fit parametrisation, the uncertainties on a $F$ are given by
\begin{equation}
\Delta F^+=\bigg[ \sum_{k=1}^n r_k^2 \, \theta \big( r_k \big) \bigg]^{1/2}, \;\;\;
\Delta F^-=\bigg[ \sum_{k=1}^n r_k^2 \, \theta \big( -r_k \big) \bigg]^{1/2}\,,
\label{eband}
\end{equation}
where $\theta$ is the Heaviside step function.
In order to propagate statistical and uncorrelated uncertainties, following Refs.~\cite{Pumplin,Stump}, we have diagonalised the covariance matrix of the best fit parameters and constructed a set of alternative parametrisations $S_{k=1..14}$ according to the standard $\Delta \chi^2=1$ criterion. The error band constructed with the help of eqs.~(\ref{eband}) and the $S_{k=1..14}$ parametrisation set is shown in 
Fig.~(\ref{xL-dependence}). The latter is narrower than individual errors on parameters obtained from fixed-$x_L$ fits. This error reduction is in fact due to the $x_L$-combination,  and can be understood considering for example $B_q$, which is just a constant as a function of $x_L$. In the combined fit, this parameter is constrained by 203 points rather than 29 of a single $x_L$ bin and so it is determined far more precisely.
Since,however, fixed-$x_L$ fits, by construction, represent the best parametrisations of the data and eqs.~(\ref{coeffAq2},\ref{coeffBq2},\ref{coeffAg2}) are interpreted as a mere interpolating tool, we require that 
the accuracy of the $x_L$-combined fit does not exceeds the one of the fixed-$x_L$ fits.
We found that a conservative $\Delta \chi^2=9$ criterion matches these requirements, 
as shown in Fig.~(\ref{xL-dependence}). We also note that, incidentally, this number is 
close to the combination penalty reported in Tab.~(\ref{combo_xl-fit}).
\begin{figure}[t]
\begin{center}
\includegraphics[width=15cm,height=15cm]{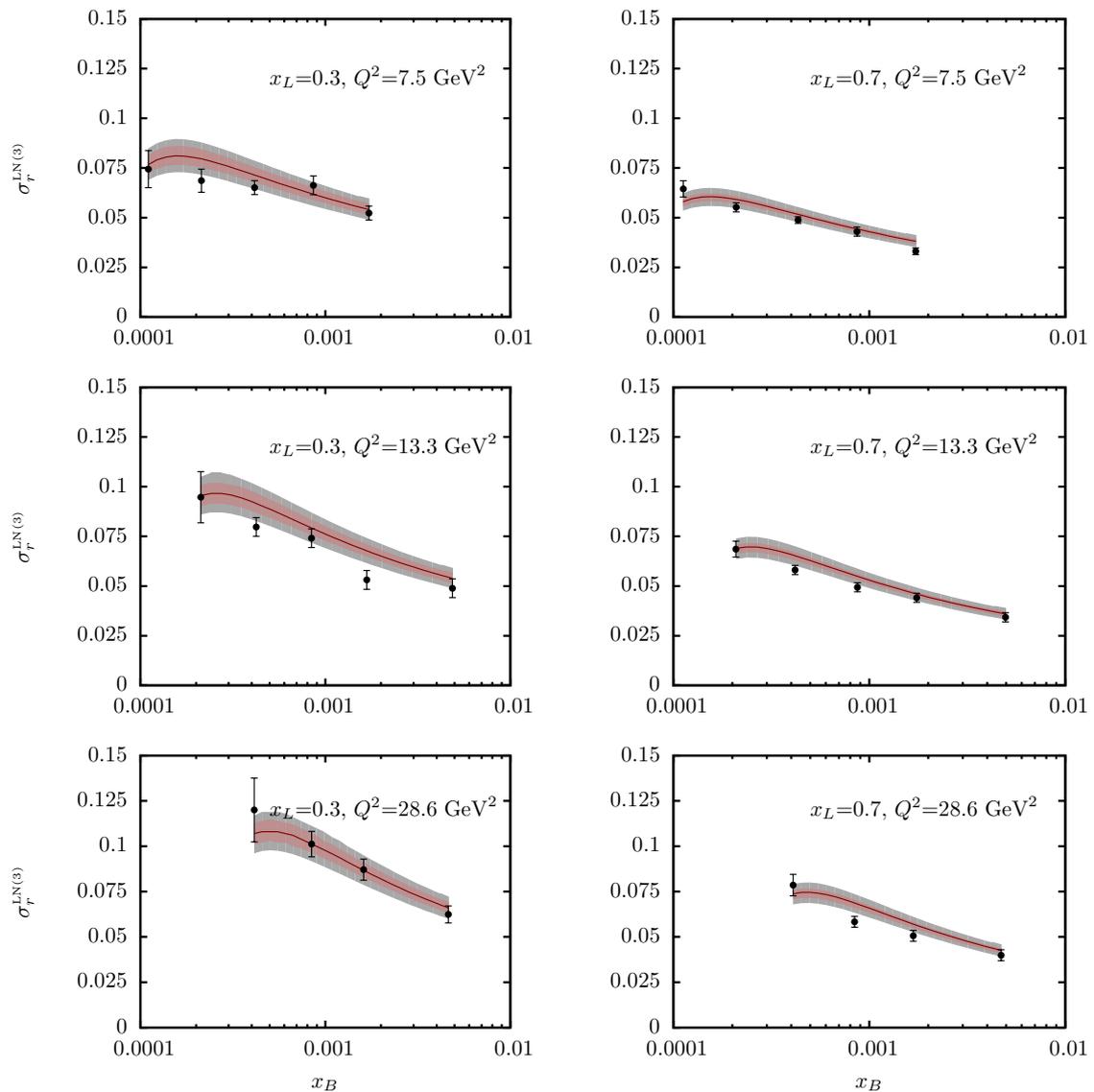}
\caption{Best fit predictions compared to ZEUS data~\cite{ZEUS_02_used}. 
The error bars associated with the data points show the sum in quadrature of the statistical and total systematic uncertainty. The light red error band corresponds to $\Delta \chi^2=9$ and it constructed with $S_{k=1..14}$.  The grey error band is constructed as the quadratic sum of statistical ($S_{k=1..14}$) and systematic ($S_{k=15..34}$) contributions.}
\label{Bestfit_vs_ZEUS}
\end{center}
\end{figure}

We now turn to the inclusion of systematics in the error analysis.
In data from Ref.~\cite{H1_10} nine systematics sources are identified plus the 
luminosity uncertainty~\cite{pc}. For each of them we performed, according to the so-called offset method, alternative fits in which each $s_k$ is held fixed in turn either to -1 or +1 and produced the parametrisation set $S_{k=15..34}$. For some sources, for example the 5\% luminosity uncertainty common to all data points, the shifts induce steady variation of the $\chi^2$ and mostly correlates with 
the central values of the normalisations coefficients $a_1$ and $a_3$. 
The impact of the propagation of systematic errors are presented  
in Fig.~(\ref{Bestfit_vs_ZEUS}) where the best fit predictions
are compared to ZEUS data~\cite{ZEUS_02_used}. The latter 
are presented in terms of reduced cross sections as a function of
$x_B$ in different bins of $x_L$ and $Q^2$ and integrated up to the 
same $p_{T,max}=0.2$ GeV, as in the H1 analysis from which nFFs are obtained.
\begin{figure}[t]
\begin{center}
\includegraphics[width=15cm,height=15cm]{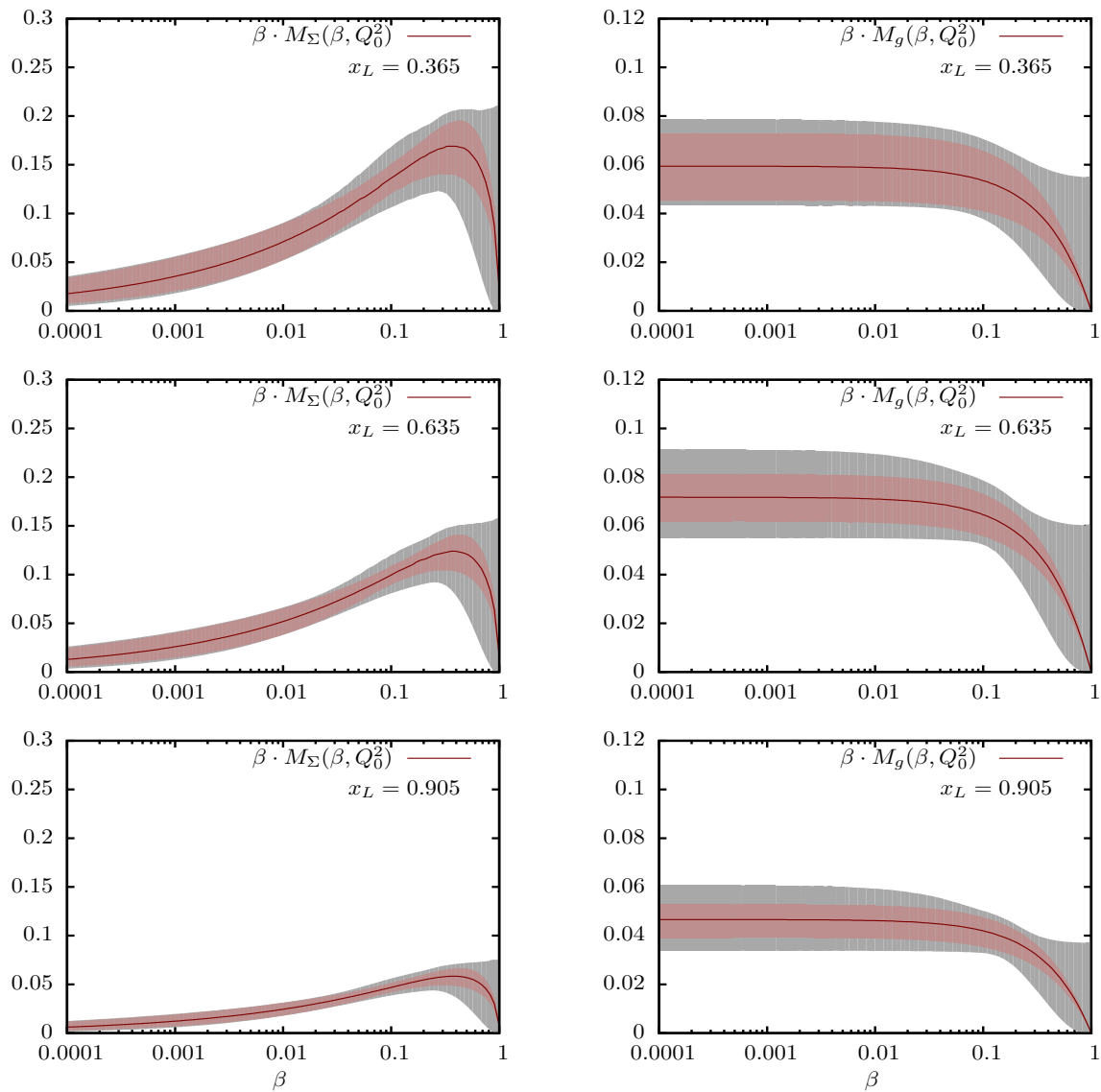}
\caption{Singlet (left) and gluon (right) momentum distributions at the initial scale 
$Q_0^2=1$ Ge$\mbox{V}^2$ in three representative bins of $x_L$. The light red band is obtained with the $\Delta \chi^2=9$ criterion, while the grey one with additional $S_{k=35..38}$ parametrisation set added in quadrature.}
\label{xparton_different_xL_at_Q0}
\end{center}
\end{figure}
The effects induced by systematic errors are significant, as shown by the light 
grey band in Fig.~(\ref{Bestfit_vs_ZEUS}). After taking into account all error sources
we find that predictions based on nFFs describe ZEUS data both in shape and 
normalisation.

We conclude this section attempting an estimate of theoretical errors. Among them we mention the ones related to pQCD settings and 
the ones related to the choice of the parametrisation of initial conditions. The latter 
are by far dominant in the present analysis so we focus on them in the following.
The impact of the functional form used for the $x_L$-combination is estimated
in 8.8 $\chi^2$ units, which corresponds to the combination penalty term. 
The uncertainties associated to the functional forms choosen for the $\beta$ dependence are not so easily quantified. 
In the following we restrict ourserlves to study of the impact 
of the main assumptions in the parametrisation of initial conditions at large-$\beta$.
The parameters controlling the large-$\beta$ behaviour, being almost unconstrained by the data, were held fixed in the fit, \textsl{i.e.} $C_q=0.5$ and $C_g=1$. This implies that the error propagation 
produces, as shown in Fig.~(\ref{xparton_different_xL_at_Q0}), an artificial shrinkage of the (light red) error band at large $\beta$ and by no means represents a correct error estimate in this limit.
\begin{figure}[t]
\begin{center}
\includegraphics[width=15cm,height=15cm]{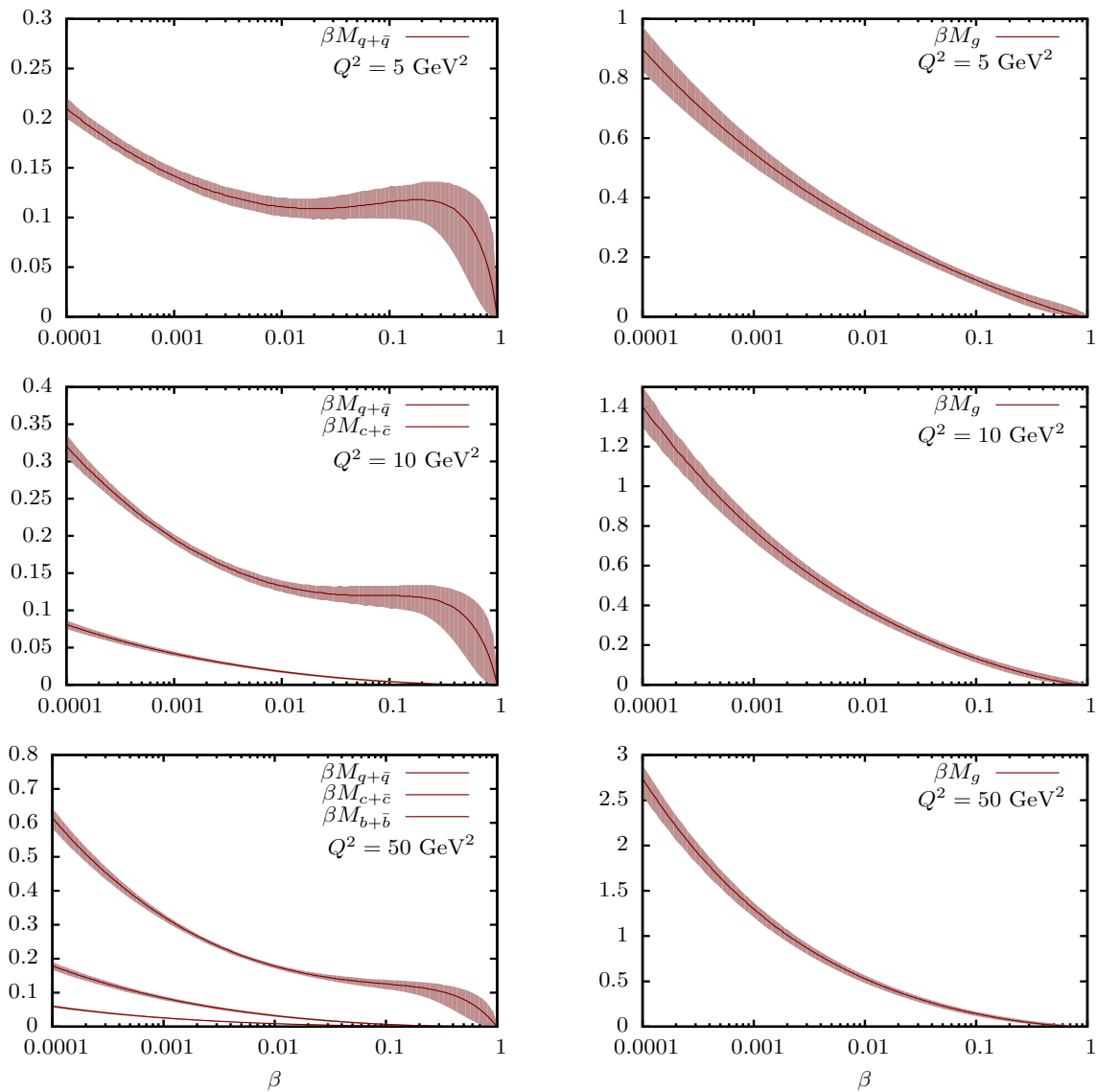}
\caption{Momentum distributions at various $Q^2$ at $x_L=0.635$. The light red band,
corresponding to the $\Delta \chi^2=9$ criterion, is obtained with the $S_{k=1..14}$ and $S_{k=35..38}$ parametrisation sets.}
\label{xparton_different_Q}
\end{center}
\end{figure}
In order to quantify the errors introduced by these assumptions, 
we performed four additional fits in which the $C_q$ and $C_g$ parameters are kept fixed 
in the minimisation but choosen in the following combinations:
$(C_q,C_g)=(0,1),(2,1),(0.5,0),(0.5,3)$. The latter values are choosen such that fits return a difference of around $\Delta \chi^2=9$ with respect to best fit. 
The latter four alternative parametrisations, $S_{k=35..38}$, are used to produce the light grey errorband shown in Fig.~(\ref{xparton_different_xL_at_Q0}) and shows
the degree of underdetermination of the distributions in this region.
These additional parametrisations can be especially useful to propagate uncertainties to observables which require nFFs large-$\beta$ extrapolation, for example jet cross section. 
Less problematic appear the $Q^2$ extrapolation, since the latter is fully predicted by the theory. In Fig.~(\ref{xparton_different_Q}) we present the initial condition 
at $x_L=0.635$ for three different values of $Q^2$.  
It is interesting to note that the uncertanties on the nFFs increase as $Q^2$ decrease. 
This effect can be partly ascribed to the fact that no data point with $Q^2$ below 7.3 Ge$\mbox{V}^2$ is included in the fit. But, more importantly, it has to be ascribed  
to QCD evolution: small displacements of the parametrisations
at high $Q^2$, where they are actually constrained by the data, turn
into large fluctuations of the initial conditions at $Q_0^2$, due to the logarithmic 
nature of QCD evolution equations.

\section{Conclusions}
\label{conclusions}
\noindent
In this paper we have presented a perturbative QCD analysis and extraction of neutron fracture functions from forward neutron production in DIS in HERA kinematics. 
Data can be decribed by the leading-twist approximation implied by the hard-scattering factorisation formula and perturbative QCD evolution down to the lowest values of $Q^2$ accessed by the experiment.
The results of the fit and within the precision of the present data, indicate that the proton-vertex factorisation hypothesis is supported to good accuracy, a fact which is likely to be related to the relative low $\beta$ regime accessed by the measuraments.
The nFFs low-$Q^2$ extrapolation, although with large uncertainties, 
can be used to address the impact of absorptive effects going from 
the DIS to the photoproduction regime.  
The predictions based on the obtained nFFs have been succesfully compared 
to ZEUS data and an error estimation on nFFs has been provided.   
The obtained nFFs parametrisation obtained in NLO QCD is a quantitative tools which can be used in factorisation tests in processes with a tagged forward neutron. In this context we mention dijet photoproduction in $ep$ collisions and
dijet or Drell-Yan pair production in hadronic collisions.
For the latter process, next-to-leading order corrections have been estimated in Ref.~\cite{aDY_ceccopieri}. The nFFs parametrisation and error set are available upon request to the author and are provided as fortran steering file for the \texttt{QCDNUM17} package.
 
\section*{Acknowledgements}
\noindent
We warmly thanks Armen Bunyatyan and Vitaly Dodonov for numerous discussions
on the measuraments and for providing us the breakdown of the systematics errors.

\end{document}